\documentclass[11pt,a4paper]{article}
\pdfoutput=1
\usepackage[UKenglish]{babel}
\usepackage[utf8]{inputenc}
\usepackage{jcappub,longtable,lscape,physics,xspace,amsmath,multirow,microtype}
\newcommand{\etal}{{et~al.}\@\xspace}
\newcommand{\eg }{{e.g.}\@\xspace}
\newcommand{\ie}{{i.e.}\@\xspace}
\newcommand{\SimProp}{\textit{SimProp}\xspace}
\newcommand{\mail}[1]{\href{mailto:#1}{\nolinkurl{#1}}}
\newcommand{\AppendixRef}{appendix~\ref}
\newcommand{\SectionRef}{section~\ref}
\newcommand{\TableRef}{table~\ref}
\newcommand{\FigureRef}{figure~\ref}
\newcommand{\ReferenceRef}{ref.~\cite}
\newcommand{\meV}{\mathrm{meV}}
\newcommand{\eV}{\mathrm{eV}}
\newcommand{\MeV}{\mathrm{MeV}}
\newcommand{\PeV}{\mathrm{PeV}}
\newcommand{\EeV}{\mathrm{EeV}}
\newcommand{\K}{\mathrm{K}}
\newcommand{\km}{\mathrm{km}}
\newcommand{\mb}{\mathrm{mb}}
\newcommand{\Mpc}{\mathrm{Mpc}}
\newcommand{\s}{\mathrm{s}}
\newcommand{\yr}{\mathrm{yr}}
\newcommand{\Einj}{E_\text{inj}}
\newcommand{\Ainj}{A_\text{inj}}
\newcommand{\zinj}{z_\text{inj}}
\newcommand{\Om}{\Omega_\text{m}}
\newcommand{\OL}{\Omega_\Lambda}
\newcommand{\kB}{k_\text{B}}
\newcommand{\p}{\mathrm{p}}
\newcommand{\n}{\mathrm{n}}
\newcommand{\e}{\mathrm{e}}

\title{\boldmath $SimProp$~v2r4: Monte Carlo simulation code for UHECR propagation}

\collaboration{The SimProp developers:}
\author[a,b]{Roberto Aloisio,}
\author[c]{Denise Boncioli,}
\author[d]{Armando di~Matteo,}
\author[b,1]{Aurelio~F.~Grillo,\note{Deceased}}
\author[a,e]{Sergio Petrera}
\author[e]{and Francesco Salamida}

\affiliation[a]{Gran Sasso Science Institute (GSSI),\\Viale Francesco Crispi 7, 67100 L'Aquila, Italy}
\affiliation[b]{INFN, Laboratori Nazionali del Gran Sasso (LNGS),\\67100 Assergi, L'Aquila, Italy}
\affiliation[c]{Deutsches Elektronen-Synchrotron (DESY),\\Platanenallee 6, 15738 Zeuthen, Germany}
\affiliation[d]{Service de Physique Th\'eorique, CP225, Universit\'e Libre de Bruxelles (ULB),\\Boulevard du Triomphe (Campus de la Plaine), 1050 Brussels, Belgium}
\affiliation[e]{Department of Physical and Chemical Sciences (DSFC), University of L'Aquila,\\Via Vetoio (Coppito 1), 67100 L'Aquila, Italy}
\emailAdd{SimProp-dev@aquila.infn.it}

\abstract{We introduce the new version of \SimProp, a Monte Carlo code for simulating the propagation of ultra-high energy cosmic rays in intergalactic space.  This version, \SimProp~v2r4, together with an overall improvement of the code capabilities with a substantial reduction in the computation time, also computes secondary cosmogenic particles such as electron--positron pairs and gamma rays produced during the propagation of ultra-high energy cosmic rays. As recently pointed out by several authors, the flux of this secondary radiation and its products, within reach of the current observatories, provides useful information about models of ultra-high energy cosmic ray sources which would be hard to discriminate otherwise.}

\keywords{ultra-high energy cosmic rays, Monte Carlo simulations}

\begin{document}
\maketitle
\flushbottom

\section{Introduction}
Cosmic rays (high-energy charged particles from outer space, mainly protons and other nuclei) with energies up to a few hundred EeV ($1~\EeV = 10^{18}~\eV \approx 0.16~\mathrm{J}$) have been observed since the 1960s.  Half a century later, the origin of such particles is still uncertain, but there is now a wide consensus that most of the cosmic rays with energies above a few EeV (known as ultra-high-energy cosmic rays, UHECRs) originate from extragalactic sources \cite{bib:AloisioTrans}. If this is the case, 
 the energy, type and trajectory of propagating particles can change in their intergalactic journey through different processes. 
These include: (i)~the adiabatic energy loss that all particles travelling cosmological distances undergo due to the expansion of the Universe (the redshift loss); (ii)~photonuclear interactions with cosmic microwave background (CMB) and infrared/visible/ultraviolet extragalactic background light (EBL) photons; and (iii)~deflections by intergalactic and galactic magnetic fields.  
Photonuclear interactions include electron--positron pair production, disintegration of nuclei (whereby one or more nucleons or other light fragments are ejected from a nucleus), and photohadronic interactions producing one or more mesons (mainly pions).
These processes need to be taken into account when predicting the energy spectrum, mass composition and distribution of arrival directions of UHECRs expected at the Earth assuming a given model for their sources. In addition, photonuclear processes may result in the production of secondary particles such as nucleons, electron--positron pairs, neutrinos and gamma rays. The detection of this radiation or lack thereof can further help in constraining scenarios for the sources of UHECRs.

Some of the quantities relevant to UHECR propagation are known with good precision. These include: (i) the redshift energy loss rate (which in the standard $\Lambda$CDM cosmology can be computed from the FLRW metric as $-\qty(\dd{\ln E}/\dd{t})_\text{ad} = H(z) = H_0 \sqrt{(1+z)^3\Om + \OL}$, where $H_0 \approx 70~\km/\s/\Mpc$, $\Om \approx 0.3$ and $\OL \approx 0.7$), (ii) the CMB spectrum and its cosmological evolution (a blackbody spectrum with temperature $T=(1+z)T_0$, $T_0\approx 2.7~\K$), (iii) photonuclear cross sections for pair production (computed from the Bethe--Heitler formula) and pion production (measured in accelerator-based experiments). Other quantities are still considerably uncertain, and phenomenological models are needed to estimate them; these include: (i) the spectrum and cosmological evolution of the EBL (especially in the far infrared and at high redshifts) and (ii) exclusive cross sections for certain photodisintegration channels (especially those where charged fragments are ejected). These uncertainties have a non-negligible impact on UHECR propagation studies as widely discussed in ref.~\cite{bib:SALpropa}.

\SimProp is a simple and fast Monte Carlo code for simulating UHECR propagation through intergalactic space. The processes taken into account are the redshift energy loss, photonuclear interactions with background photons, and decays of unstable particles (pions, muons and unstable nuclei).
The photon backgrounds taken into account in \SimProp are the CMB and the EBL, with several models available for the EBL spectrum and evolution (\SectionRef{sec:EBL}).
The photonuclear interaction processes implemented in \SimProp are
the electron--positron pair production, approximated as a continuous energy loss process;
the disintegration of nuclei, for which different models are available (\SectionRef{sec:disi}); and
photohadronic interactions, all approximated as single-pion production with branching ratios computed assuming isospin invariance, distribution of outgoing pion directions taken to be isotropic in the centre-of-mass (CoM) frame, and nuclei treated as collections of free nucleons. 
Particle decays are approximated as instantaneous, as decay lengths are generally much shorter than all
other relevant length scales.
Intergalactic magnetic fields are neglected in \SimProp; consequently, UHECR propagation through intergalactic space is taken to be rectilinear.

The main new feature of \SimProp~v2r4 compared to the previous versions \cite{bib:SPv2r0,bib:SPv2r1,bib:SPv2r2,bib:SPv2r3} is the computation of the production rates of high-energy secondary electrons, positrons and gamma rays produced during the propagation of UHECRs. Such particles can initiate electromagnetic cascades via inverse Compton scattering and electron--positron pair production against photons of the CMB, EBL and radio backgrounds, resulting in a diffuse background of photons up to TeV energies. Measurements of the diffuse extragalactic gamma-ray background can provide an upper bound to the energy density of such cascades, which can be used to constrain models of UHECR sources \cite{Ferrigno:2004am,Berezinsky:2010xa,Liu:2016brs,Berezinsky:2016jys}.

This paper is structured as follows: in \SectionRef{sec:history} we briefly recall the history of \SimProp versions; in \SectionRef{sec:code} we describe technical details of the implementation of \SimProp~v2r4; in \SectionRef{sec:models} we list and discuss the various models of extragalactic background light and photodisintegration cross sections available; in \SectionRef{sec:appli} we discuss the new capabilities of \SimProp~v2r4 and how they can be useful for testing UHECR source models. In \AppendixRef{app:inout} we describe the command-line options \SimProp~v2r4 accepts and the structure of its output files. In \AppendixRef{app:distances} we define the various measures of cosmological distances used in the text. Finally, in \AppendixRef{app:time} we report the computation time required by \SimProp simulations in a few selected situations.

Unless otherwise specified, natural units $c = \hbar = 1$ are used.

\SimProp is available upon request to \mail{SimProp-dev@aquila.infn.it}.

\section{Motivation and history}\label{sec:history}
\SimProp was originally developed as a refinement of the analytic models by Aloisio, Berezinsky and
Grigorieva \cite{bib:Aloisio1,bib:Aloisio2} in order to have a publicly available
Monte Carlo code for the community to use, at a time when most UHECR propagation studies
used closed-source simulation codes such as that by Allard \etal \cite{bib:Allard} or Hooper \etal \cite{bib:Hooper}.

Since the first \SimProp release \cite{bib:SPv2r0}, more public codes have become available, including CRPropa \cite{bib:CRP2,bib:CRP3}, Hermes \cite{DeDomenico:2013psa}, and TransportCR \cite{Kalashev:2014xna}. The usefulness of having several independently developed codes is that it allows us to cross-check results, improving the computation reliability.  For example, through the comparison of the results obtained with CRPropa and \SimProp it was possible to assess the relevance of uncertainties in a few astrophysical and nuclear physics quantities \cite{bib:SALpropa}, neglected before. The same comparison brought to the conclusion that CRPropa and \SimProp show a substantial agreement \cite{bib:SALpropa}, confirming the validity of these computation schemes. 

Another test of the correctness of \SimProp was the comparison of Engel \etal \cite{Engel:2001hd} neutrino fluxes to \SimProp results in the same physical scenario\footnote{\label{fn:1}Note that the model considered there assumes a pure proton injection with practically no injection cutoff, which is no longer considered plausible due to UHECR composition data from the Pierre Auger Observatory \cite{Aab:2014aea}, and neglects interactions with the EBL, which are the dominant origin of cosmogenic neutrinos below about 0.1~EeV \cite{bib:SALpropa}.} showing very good agreement except in the electron antineutrino flux at the highest energies (which is in any case subdominant) due to the approximation of all photohadronic interactions as single-pion production in \SimProp \cite{bib:SPv2r2}. Finally, a new systematic comparison of the outputs of \SimProp and CRPropa mainly in terms of secondary neutrinos and gamma rays is planned and will be soon released \cite{bib:newSAL}.

In the first released version, \SimProp~v2r0~\cite{bib:SPv2r0}, the only processes treated stochastically were
the sampling of the injection redshift, the energy of primary particles
and the photodisintegration of nuclei.
All other processes, namely the adiabatic energy loss, pair production and pion production, were
treated deterministically, as in the analytic models \SimProp was based on \cite{bib:Aloisio1,bib:Aloisio2}.
Pion production was only taken into account for protons interacting with CMB photons and approximated as a continuous energy loss.
Photodisintegration was treated according to the Puget--Stecker--Bredekamp (PSB) model~\cite{bib:PSBsigma}
as refined by Stecker and Salamon~\cite{bib:SSthresholds}, taking into account
the EBL using the Stecker \etal model~\cite{bib:SteckerEBL} or a power-law approximation thereof
as well as the CMB. 

Starting from the following version, \SimProp~v2r1~\cite{bib:SPv2r1}, the pion production process is also treated stochastically, for both free protons and nucleons bound within nuclei, in order to compute fluxes of $\EeV$ secondary neutrinos thereby produced, also assessing the accuracy of the continuous energy loss approximation for that process~\cite{bib:SALpropa}.

\SimProp~v2r2~\cite{bib:SPv2r2}, apart from fixing a few bugs in \SimProp~v2r1, added an option to also take into account pion production on the EBL, in order to compute secondary neutrino fluxes down to $\PeV$~energies \cite{Aloisio:2015ega}, also including the choice of using the Kneiske \etal model~\cite{bib:KneiskeEBL} for the spectrum and cosmological evolution of the EBL.

Starting from \SimProp~v2r3~\cite{bib:SPv2r3,bib:AdMthesis}, the user can also choose the EBL models from Do\-m\'in\-guez \etal \cite{bib:DominguezEBL} (best-fit, lower-limit or upper-limit) or Gilmore \etal \cite{bib:GilmoreEBL} (fiducial), and one of four different parametrizations for photodisintegration cross sections with user-defined parameter values. Processes where alpha particles are ejected (\eg ${^{A}Z} + \gamma \to {^{A-4}(Z-2)} + {^4\mathrm{He}}$) are also implemented, and their cross sections can be scaled by a user-defined factor via a simple command-line option, allowing the user to assess the effects of these poorly known quantities on the results. 

In the present version, \SimProp~v2r4, secondary electron--positron pairs and photons produced by the propagation of UHECRs can also be written in the output file, so that external tools, such as  DINT~\cite{bib:DINT}, ELMAG~\cite{bib:ELMAG}, or $\mathit{EleCa}$~\cite{bib:EleCa} can then be used to simulate the electromagnetic cascades initiated by such particles interacting with CMB, EBL and radio backgrounds. In addition, the user can now specify an arbitrary spectral index for the energy distribution of the injected particles. Finally, a few adjustments and bug fixes have been applied to the new version of \SimProp, reducing the overall computation time; a few examples of the computation times required by the current \SimProp version are reported in \AppendixRef{app:time}.

\section{Structure of the code}\label{sec:code}
\SimProp is written in C++, and makes use of features from the ROOT~\cite{bib:ROOT} framework. The random number generator used in \SimProp is {\tt TRandom3} from ROOT, which is based on the Mersenne twister~\cite{bib:Mersennetwister}.

A $\SimProp$ run consists of $N$ events, each including the generation of a primary particle with mass number~$A_\text{inj}$, initial energy~$\Einj$  sampled from a power-law distribution from~$E_{\min}$ to~$E_{\max}$ with spectral index $\gamma$ (\ie $\dd{N} \propto \Einj^{-\gamma} \dd{\Einj}$), and source redshift~$\zinj$ uniformly distributed from~$z_{\min}$ to~$z_{\max}$.\footnote{This can be converted to different distributions by weighing each event by an appropriate function of $\zinj$, for example by $w(\zinj)\propto{1}/{\sqrt{(1+\zinj)^3\Om+\OL}}$ for a constant density of sources per unit comoving volume (see \AppendixRef{app:distances}).} The propagation of particles is followed, along with that of any secondary particles produced during propagation. The injection parameters $N$, $A_\text{inj}$, $E_{\min}$, $E_{\max}$, $\gamma$, $z_{\min}$ and~$z_{\max}$ can be specified via command-line options (\AppendixRef{app:inout}).

The interval between the production of a particle (at injection for primaries, or at an interaction for secondaries) and its decay, stochastic interaction or arrival at the Earth is called a branch. During a branch, a particle keeps the same mass number and electric charge, but loses energy through continuous processes, such as adiabatic and pair production energy loss. Each of the outgoing particles in a stochastic interaction or decay starts a new branch, even if it is of the same type as the parent (\eg  as in neutral pion production). Secondary electrons and positrons from pair production also start a new branch, but the parent nucleus continues the same branch.

\subsection{Propagation of protons and stable nuclei}
When propagating a proton or a stable nucleus, the redshift interval between its production point $z_\text{prod}$ and $0$ (Earth) is divided into steps $z_0 = z_\text{prod}$, $z_1$, \ldots, $z_n = 0$. During each step, there are two types of processes the particle can undergo:
\begin{itemize}
\item those treated in the approximation of continuous (deterministic) energy losses, namely the adiabatic energy loss, pair production on CMB photons, and (if the option {\tt -S~-1} or {\tt -S~0} is used) pion production on CMB photons;
\item those treated as discrete (stochastic) interactions, with the interaction point and the energy, type and/or number of outgoing particles to be sampled stochastically, namely the photodisintegration of nuclei and pion production (on CMB photons by default, if the option {\tt -S~1} is used, or also on EBL photons if {\tt -S~2} is used).
\end{itemize}

\subsubsection{Continuous energy losses}
At each redshift step $(z_{i-1},z_i]$,  the processes computed in the continuous energy loss approximation are treated by numerically integrating the differential equation for the logarithm of the Lorentz factor $\ln \Gamma$ from $z_{i-1}$ to $z_{i}$, as described in ref.~\cite{bib:SPv2r0}. The values used for the cosmological parameters are  $H_0 = 7.17\times 10^{-11}~\yr^{-1}\approx 70~\km/\s/\Mpc$, $\Om = 0.3$ and $\OL = 0.7$. The pair-production energy loss rates for protons at $z=0$ are interpolated from a table computed as described in \ReferenceRef{Berezinsky:2002nc}. For nuclei heavier than protons,  pair production energy losses are approximated as those of protons multiplied by the factor~$A/4$ (which is typically equal or very close to the exact factor~$Z^2/A$).

The number of secondary electron--positron pairs generated in the step and optionally written to the output file is sampled from a Poisson distribution with mean \begin{align}n_\text{pairs} &= \frac{1}{K} \ln (\frac{\Gamma_{i-1}}{1+z_{i-1}}\frac{1+z_i}{\Gamma_i}), & \text{where~}K&=\frac{2m_\e}{Am_N + 2m_\e},\end{align} on the assumption that each pair is generated with energy $K$~times that of the parent particle, as it is at the kinematic threshold. The production points of the secondary pairs are chosen at random uniformly from $z_{i-1}$ to $z_i$, and the energies of the electrons and positrons as $K Am_N\Gamma/2$, where $\log\Gamma$ is chosen at random uniformly from $\log\Gamma_{i-1}$ to $\log\Gamma_i$. 

The approximation that the fraction of energy carried by each pair is $K$ is not very accurate for energies much higher than the threshold. Nevertheless, high-energy electrons produced at even moderate~$z$ result in cascading photons at TeV energies. This radiation has a spectrum quite independent of the individual energies of the initial electrons \cite{bib:secondarygamma}; the only quantity with observable consequences is the total energy emitted in pairs at each interval, which is not affected by the approximation above.

\subsubsection{Discrete interactions}
An extension of the scheme described in \ReferenceRef{bib:SPv2r0} is used in order to determine whether a particle undergoes a discrete interaction within a given redshift step and, if it does, to sample the interaction point within the step, type, number and energies of the interaction products.

The total interaction rate (probability per unit time) $1/\tau$ of an ultrarelativistic particle with background photons is given by \begin{equation}
\frac{1}{\tau} = \frac{1}{2\Gamma^2}\int_{\epsilon'_{\min}}^{2\Gamma\epsilon}\int_{\epsilon=0}^{+\infty}\frac{n_\gamma(\epsilon)}{\epsilon^2}\dd{\epsilon} \sigma(\epsilon') \epsilon' \dd{\epsilon'} \label{eq:intrate},
\end{equation} 
where $\epsilon'_{\min}$ is the threshold energy of the process at hand, $n_\gamma(\epsilon)$ is the density of background photons, $\sigma(\epsilon')$ is the interaction cross section, $\Gamma$ is the particle's Lorentz factor and all primed quantities are intended in the CoM reference frame. 
Since $1/\tau$ is linear in both the background photon density and the cross sections, it can be computed as the sum of the interaction rates for each type of process and photon background, $\tau^{-1}=\sum_{ij}\tau_{ij}^{-1}$. When the particle is found to have interacted, the probability that the interaction was of type~$i$ with a photon of the $j$-th background is then $p_{ij}=\tau_{ij}^{-1}/\tau^{-1}$.  

If we introduce the quantities
\begin{align} 
  I(\epsilon) &= \int_\epsilon^{+\infty}  \frac{n_\gamma(\varepsilon)}{2\varepsilon^2} d\varepsilon, &
  \Phi(s) &= \int_{s_{\min}}^{s} (s'-m^2)\sigma(s')\,ds'
          = 4m^2\int_{\epsilon'_{\min}}^{\epsilon'} \varepsilon'\sigma(\varepsilon')\,d\varepsilon' \label{eq:Phi},
\end{align} where $m$ is the mass of the particle (a nucleus in the case of disintegration and a nucleon in the case of pion production) and $s$ is the CoM energy squared $s=m^2+2m\epsilon'$, then eq.~\eqref{eq:intrate} can be rewritten as 
\begin{align}
\frac{1}{\tau} = \frac{1}{4m^2\Gamma^2} \int_{{\epsilon'_{\min}}/{2\Gamma}}^{+\infty} \Phi(m^2+4m\Gamma\epsilon) \frac{n_\gamma(\epsilon)}{2\epsilon^2} \dd{\epsilon} 
= \frac{1}{\Gamma^2} \int_{\epsilon'_{\min}}^{+\infty} I \left(\frac{\epsilon'}{2\Gamma}\right) \epsilon'\sigma(\epsilon') \dd{\epsilon'}. \label{tau}
\end{align}

The photon background $n_{\gamma}$ is the sum of two terms, one for the CMB and one for the EBL. The CMB spectrum is precisely known at all redshifts, being a blackbody spectrum: 
\begin{align}
n_\text{CMB}(\epsilon) &= \frac{1}{\pi^2} \frac{\epsilon^2}{\exp(\epsilon/\kB T)-1}, \label{eq:CMB} & I_\text{CMB}(\epsilon) &= -\frac{\kB T}{2\pi^2}\ln(1-\exp(-\frac{\epsilon}{\kB T})),
\end{align}
where $T = (1+z)T_0$, and the value of $\kB T_0$ used in \SimProp is $0.2327~\meV$ ($T_0 \approx 2.7~\K$).
This implies that the $z$-dependence of the interaction lengths on the CMB is given by $\tau_\text{CMB}^{-1}(\Gamma,z) = (1+z)^3 \tau^{-1}_\text{CMB}((1+z)\Gamma,z=0)$. On the other hand, EBL is not precisely known, and needs to be approximated using phenomenological models; those available in \SimProp are listed in \SectionRef{sec:EBL}.

As for the cross sections, we used the total photohadronic cross section computed by SOPHIA~\cite{bib:SOPHIA} for protons as the pion photoproduction cross section~$\sigma_\text{pion}(\epsilon')$, and numerically integrated the second integral in eq.~\eqref{eq:Phi} to obtain a table of values from which we interpolate $\Phi_\text{pion}(s)$. As for photodisintegration, various models are available, listed in \SectionRef{sec:disi}.

From these quantities, $\tau_\text{pion,CMB}^{-1}(A,\Gamma,z)$ is computed as
\begin{equation}
\tau_\text{pion,CMB}^{-1}(A,\Gamma,z)=(1+z)^3A \tau^{-1}_\text{pion,CMB}(\text{proton},(1+z)\Gamma,z=0),
\end{equation} where $\tau^{-1}_\text{pion,CMB}$~for protons at~$z=0$ is interpolated from a table with values obtained by numerically integrating the first integral in eq.~\eqref{tau}.  $\tau^{-1}_\text{pion,EBL}$ is computed as a function of $\Gamma$ and $z$ via a 2D interpolation from a table obtained by numerically integrating the second integral in eq.~\eqref{tau}. As for pion production, we assume that a nucleus behaves as $A$ independent nucleons, \ie, $\tau_\text{pion}^{-1}(A,\Gamma,z) = A\tau_\text{pion}^{-1}(\text{proton},\Gamma,z)$ because the energies involved are much larger than the binding energy per nucleon. Finally, if needed, $\tau^{-1}_\text{disi}$ is computed by numerically integrating the second integral in eq.~\eqref{tau}. Then we have $\tau^{-1} = \tau_\text{pion,CMB}^{-1} + \tau_\text{pion,EBL}^{-1} + \tau_\text{disi}^{-1}$. 

After a particle is found to have interacted and the type of interaction is determined, the number, type and energy of the outgoing particles are sampled as follows.

\paragraph{Photodisintegration.}
When a nucleus is photodisintegrated, one of the channels implemented in the photodisintegration model used (see \SectionRef{sec:disi}) is chosen at random, with probabilities proportional to their respective interaction rates. The energy of the parent nucleus is assumed to be split among the residual nucleus and the ejected fragments proportionally to their mass, \ie all the fragments inherit the Lorentz factor of the original nucleus.

\paragraph{Pion photoproduction.}
When a pion is photoproduced, if the incoming particle is a nucleus, the nucleon that undergoes the interaction is chosen at random. We approximate all photohadronic processes as single-pion production; assuming isospin invariance, a neutral pion is produced ($\p + \gamma \to \p + \pi^0$, $\n + \gamma \to \n + \pi^0$) with probability~$2/3$ and a charged pion is produced ($\p + \gamma \to \n + \pi^+$, $\n + \gamma \to \p + \pi^-$)  with probability~$1/3$.

In order to sample the pion energy, first the photon energy $\epsilon$ in the lab frame is sampled from its marginal distribution\footnote{In practice, we use the fact that the marginal distribution of $\epsilon$ corresponds to a distribution of $I(\epsilon)$ proportional to $\Phi(m^2+4m\Gamma\epsilon)$, and the conditional distribution of $s$ given $\epsilon$ corresponds to a uniform distribution of $\Phi(s)$, so we actually sample $I$ and $\Phi$ and invert the functions to find the corresponding $\epsilon$ and $s$.}  
\begin{align}
p(\epsilon)\dd{\epsilon} &= \frac{\tau}{4m^2\Gamma^2}\Phi(m^2+4m\Gamma\epsilon)\frac{n_{\gamma}(\epsilon)}{2\epsilon^2}\dd{\epsilon}, \quad \frac{\epsilon'_{\min}}{2\Gamma} <\epsilon < +\infty,
\end{align} 
where $m$~is the nucleon mass and~$\epsilon'_{\min} = m_{\pi} + m_{\pi}^2/2m$; then, given~$\epsilon$, the squared CoM energy~$s$ is sampled from its conditional distribution 
\begin{align}
p(s|\epsilon)\dd{s} &= \frac{(s-m^2)\sigma(s)\dd{s}}{\Phi(m^2+4m\Gamma\epsilon)}, \quad (m+m_{\pi})^2 < s < m^2+4m\Gamma\epsilon~,
\end{align} from~$s$  the pion energy and momentum in the CoM frame are calculated as  
\begin{align}
E^*_{\pi} &= \frac{s - m^2 + m_{\pi}^2}{2\sqrt{s}}; & p^*_{\pi} &= \frac{\sqrt{\left(s-(m+m_{\pi})^2\right)\left(s-(m-m_{\pi})^2\right)}}{2\sqrt{s}}
\end{align}
and the Lorentz factor of the transformation from the CoM frame to the lab frame as $\gamma = m\Gamma/\sqrt{s}$; then the pion energy is converted to the lab frame as~$E_{\pi} = \gamma(E^*_{\pi}+p^*_{\pi}\cos\theta_{\pi})$, where the distribution of~$\theta_{\pi}$ is approximated as isotropic ($\cos\theta_{\pi}$ uniformly distributed from~$-1$ to~$1$). In the lab frame, the momentum component orthogonal to the original direction is much smaller than that parallel to it, by a factor of order $\epsilon/E \sim 10^{-20}$, so we neglect transverse components assuming one dimensional propagation.

The outgoing particles are then the pion with energy $E_\pi$, the nucleon with energy $m\Gamma - E_\pi$, and (in the case of nuclei) the residual nucleus with mass number $A-1$ and energy $(A-1)m\Gamma$.

\subsection{Decay of unstable particles}\label{sec:dec}
When an unstable particle is produced, it is assumed to decay instantaneously, as decay lengths are generally much shorter than all other relevant length scales \cite{bib:Aloisio1,bib:Aloisio2}. The energies of the decay products are sampled as follows.

\paragraph{Beta decay of neutrons and unstable nuclei.}
Neutrons and nuclei not in the list of beta-decay stable isobars are assumed to immediately undergo beta decay. The $Q$-value of the reaction is read from a table taken from \ReferenceRef{bib:masses} or, for nuclei not on that table, estimated via the  semi-empirical mass formula.

Then, the electron energy in the nucleus rest frame $E^*_\e$ is sampled from a distribution $\propto (E^{*2}_\e-m_\e^2)^{1/2} E^*_\e (Q - (E^*_\e - m_\e))^2$ (\ie, neglecting electromagnetic effects) and the neutrino energy is calculated as $E^*_{\nu} = Q - (E^*_\e - m_\e)$; the recoil of the nucleus is neglected. The electron and neutrino energies are converted to the lab frame by $E_{\e} = \Gamma E^*_{\e} (1 - \cos\theta_\e)$ and $E_{\nu} = \Gamma E^*_{\nu} (1 - \cos\theta_\nu)$, where $\Gamma$ is the Lorentz factor of the nucleus, and $(1-\cos\theta_\e)$ and $(1-\cos\theta_\nu)$ are sampled from the uniform distribution from~$0$ to~$2$. 

The decay products are then the daughter nucleus (with the same energy and mass number $A$ as the parent, with electric charge $Z$ incremented in $\beta^-$ decay and decremented in $\beta^+$~decay), the neutrino ($\bar{\nu}_e$ in~$\beta^-$ decay, $\nu_e$ in~$\beta^+$ decay) and the electron or positron.

\paragraph{Neutral pion decay.}
A $\pi^0$ with energy~$E_\pi$ decays into two photons with energy $E_{\gamma_1}$ distributed uniformly from~0 to~$E_\pi$ and $E_{\gamma_2} = E_\pi - E_{\gamma_1}$.

\paragraph{Charged pion decay.}
A $\pi^\pm$ with energy~$E_\pi$ decays into a muon with energy~$E_\mu$ distributed uniformly from~$0$ to~$ (1-m_\mu^2/m_\pi^2)E_\pi$ and a neutrino with energy $E_\nu = E_\pi - E_\mu$.

\paragraph{Muon decay.}
A muon with energy $E_\mu$ decays into two neutrinos and an electron; their energies are sampled as follows: \begin{itemize}
\newcommand{\pone}{p^*_{\nu_1}}
\newcommand{\pones}{p^{*2}_{\nu_1}}
\newcommand{\ptwo}{p^*_{\nu_2}}
\newcommand{\ptwos}{p^{*2}_{\nu_2}}
\newcommand{\pthr}{p^*_\mathrm{e}}
\newcommand{\pthrs}{p^{*2}_\mathrm{e}}
\item the energies of the neutrinos in the muon rest frame $E^*_{\nu_1}$ and $E^*_{\nu_2}$   are sampled independently uniformly from 0 to $m_\mu/2 - m_\e^2/2m_\mu$, and that of the electron is $E^*_\e = m_\mu - E^*_{\nu_1} - E^*_{\nu_2}$;
\item the corresponding momenta are computed as $p^*_{\nu_1} = E^*_{\nu_1}$, $p^*_{\nu_2} = E^*_{\nu_2}$, $p^*_\mathrm{e} = \sqrt{E^{*2}_\mathrm{e}-m_\mathrm{e}^2}$;
\item if these values violate any of the constraints $E^*_\mathrm{e} \ge m_\mathrm{e}$, $\pone \le \ptwo + \pthr$, $\ptwo \le \pthr + \pone$, or $\pthr \le \pone + \ptwo$, they are discarded and a new $E^*_{\nu_1}, E^*_{\nu_2}$ pair is sampled;
\item the angle $\theta_{12}$ between the two neutrinos is given by \begin{equation}\cos\theta_{12} = \frac{\pthrs-\pones-\ptwos}{2\pone\ptwo};\end{equation}
\item the angle $\theta_1$ between the first neutrino and the line of sight is isotropic, \ie $\cos\theta_1$ uniform from $-1$ to $1$;
\item the angle $\phi$ between the second neutrino and the plane containing the line of sight and the first neutrino is uniform from 0 to $2\pi$;
\item the angle $\theta_2$ between the second neutrino and the line of sight is given by $\cos\theta_2 = \cos\theta_{12}\cos\theta_1 - \sin\theta_{12}\sin\theta_1\cos\phi$;
\item finally, the neutrino energies are transformed to the lab frame via $E_{\nu_1} = \gamma(E^*_{\nu_1}+p^*_{\nu_1}\cos\theta_1)$ and $E_{\nu_2} = \gamma(E^*_{\nu_2}+p^*_{\nu_2}\cos\theta_2)$, and the electron energy is computed as $E_\e = E_\mu - E_{\nu_1} - E_{\nu_2}$.
\end{itemize}

\subsection{Other particles: neutrinos, photons and electron--positron pairs}
The propagation of neutrinos is trivial: no interaction is possible and the only energy loss is the redshift loss, so a neutrino produced with energy~$E_\text{prod}$ at redshift~$z_\text{prod}$ will reach Earth with energy~$E_\text{Earth}=E_\text{prod}/(1+z_\text{prod})$.  Neutrino flavours are recorded only at the production; flavour oscillations are not implemented in the code, as it is trivially easy to calculate flavour transformations from the source to the observer\footnote{For example, assuming tribimaximal mixing, a pure $\nu_\e$~flux at production results in $\nu_\e : \nu_\mu : \nu_\tau = 5:2:2$ ratios at Earth, a pure $\nu_\mu$~or $\nu_\tau$~flux at production results in $4:7:7$~ratios at Earth, and $1:2:0$~ratios at production result in $1:1:1$ ratios at Earth.}.

The energy and production redshift of secondary photons or electron--positron pairs are recorded in the output files so that the electromagnetic cascades initiated by them can be simulated with external tools such as DINT~\cite{bib:DINT}, ELMAG~\cite{bib:ELMAG}, or $\mathit{EleCa}$~\cite{bib:EleCa}, as the propagation of photons and electron--positron pairs is not yet implemented in {\it SimProp}. Since the spectral shape of electromagnetic cascades at the Earth does not depend on the initial energy (or initiating particle type) and depends only weakly on the initial redshift \cite{bib:secondarygamma}, the expected flux of secondary gamma rays observed at the Earth can be estimated from the total energy deposited in photons and pairs in each redshift interval.

\section{Models of physical quantities}\label{sec:models}
\subsection{Extragalactic background light}\label{sec:EBL}
It is very difficult to directly measure the EBL because the foreground, the
zodiacal light, is orders of magnitude larger than the EBL itself. Therefore,
phenomenological models are used to describe its spectrum and its evolution
with redshift, based on the time evolution of galaxy populations as estimated
using different approaches. The results of these methods are not always
compatible each other, even in the case of recent models, especially for
long wavelengths (far IR) and at large redshifts.

\begin{figure}[t]
    \centering
    \includegraphics[width=0.45\textwidth]{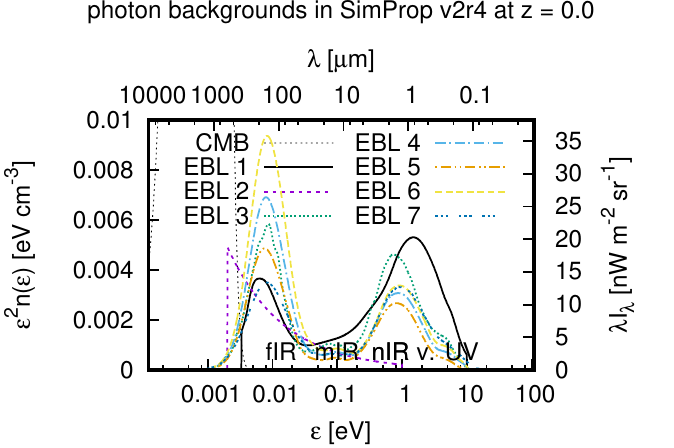}
    \includegraphics[width=0.45\textwidth]{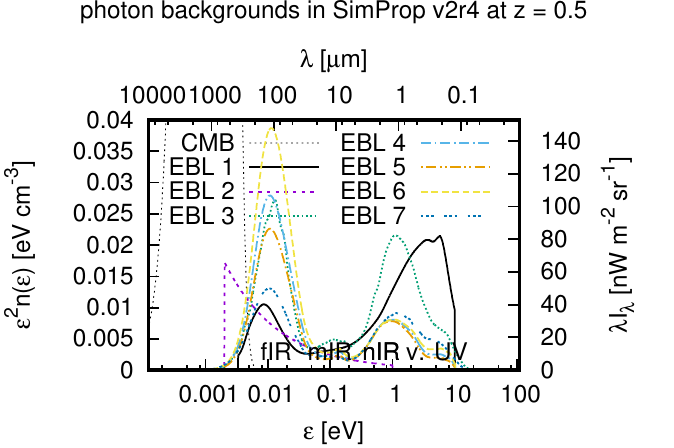}
    \includegraphics[width=0.45\textwidth]{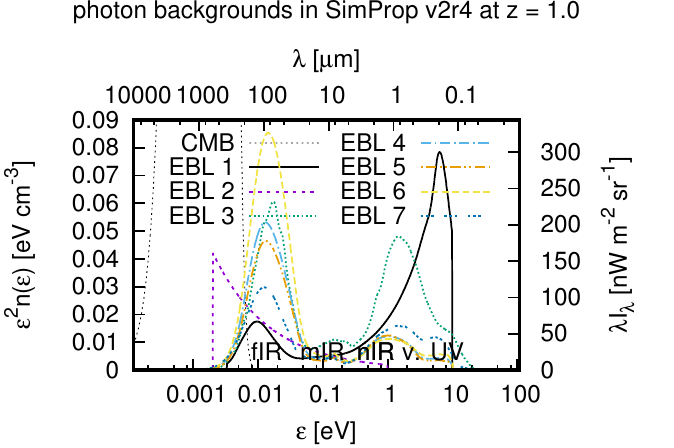}
    \caption{The EBL models implemented in \SimProp~v2r4 at $z=0.0, 0.5, 1.0$. The tails of the CMB are also shown for comparison. These models are not necessarily identical on those from the works they are based on due to the ranges and binnings of $\epsilon$ and $z$.}
    \label{fig:ebl}
\end{figure}

The following EBL models are available in \SimProp, and can be chosen via the {\tt -L} command-line option:
\begin{center}\begin{tabular}{rllll}
  No. & model & ref. & $\epsilon_{\min}$ & $\epsilon_{\max}$ \\
  \hline
  0 & none (CMB only) & ~ & ~ & ~ \\
  \textbf{1} & Stecker \etal 2006 \textbf{(default)} & \cite{bib:SteckerEBL} & $3.3~\meV$ & $\phantom{0}10~\eV$ \\
  2 & power-law approximation & ~ & $2~\meV$ & $\phantom{00}1~\eV$ \\
  3 & Kneiske \etal 2004 & \cite{bib:KneiskeEBL} & $1.26~\meV$ & $\phantom{0}11.5~\eV$ \\
  4 & Dom\'inguez \etal 2011 best fit & \cite{bib:DominguezEBL} & $1.24~\meV$ & $\phantom{0}10.2~\eV$ \\
  5 & Dom\'inguez \etal 2011 lower limit & \cite{bib:DominguezEBL} & $1.24~\meV$ & $\phantom{0}10.2~\eV$ \\
  6 & Dom\'inguez \etal 2011 upper limit & \cite{bib:DominguezEBL} & $1.24~\meV$ & $\phantom{0}10.2~\eV$ \\
  7 & Gilmore \etal 2012 fiducial & \cite{bib:GilmoreEBL} & $0.12~\meV$ & $108~\eV$ \\
\end{tabular}\end{center}
The corresponding spectral energy densities are plotted in figure~\ref{fig:ebl}. Note that EBL models up to {\tt -L 3} are included for
backward compatibility, but they are in tension with Fermi gamma-
ray observations of blazars \cite{Ackermann:2012sza}, and their use is not
recommended except for testing purposes.

In models 1, 4, 5, 6 and 7, $n_\text{EBL}(\epsilon, z)$ is found via cubic spline interpolation on a 2D grid as a function of $\epsilon$ and $z$. In model 2, it is calculated analytically as $n_\text{EBL}(\epsilon, z)=N_0 (1+z)^{3.1} \epsilon^{-2.5}$ for $z\le 1.4$ and $N_0 (1+1.4)^{3.1} \epsilon^{-2.5}$ for larger $z$.
In model 3, it is calculated as $n_\text{EBL}(\epsilon, z)=S(z)n_0(\epsilon/(1+z))$, with both factors found via cubic spline interpolation on 1D grids as functions of $z$ and $\epsilon/(1+z)$ respectively.
\subsection{Photodisintegration cross sections}\label{sec:disi}
Measurements of photodisintegration cross sections are not available for all nuclides, and when there are, sometimes only the total and/or the one-neutron ejection cross sections have been measured. Cross sections for exclusive channels in which charged fragments are ejected are hard to measure, because such ejectiles tend to undergo multiple scattering in the target. Various phenomenological models have been used in UHECR propagation studies to treat these processes, with sizeable differences among them \cite{Boncioli:2016lkt}.

The following photodisintegration models are available in \SimProp, and can be chosen via the {\tt -M} command-line option:
\begin{center}\begin{tabular}{rlll}
  No. & channels for $\epsilon' < \epsilon_1$ & shape & parameter values \\
  \hline
  \textbf{0} & one- and two-nucleon \textbf{(default)} & Gaussian & PSB with SS thresholds \\
  1 & one- and two-nucleon & Gaussian & from standard input \\
  2 & one- and two-nucleon & Lorentzian & from standard input \\
  3 & one-nucleon and alpha-particle & double Lorentzian & from standard input \\
  4 & one-nucleon and alpha-particle & Gaussian & from standard input \\
\end{tabular}\end{center}

In photodisintegration models {\tt -M~0}, {\tt -M~1} and {\tt -M~2}, the processes considered are single nucleon ejection, double nucleon ejection, and multiple nucleon ejection; the type of each nucleon ejected is sampled at random (\ie, if a nucleus with 26 protons and 30 neutrons loses a nucleon, it is assumed to be a proton with probability~$26/56$ and a neutron with probability~$30/56$); this simplifying assumption is only approximately realistic (as in reality interaction channels yielding stable nuclei are more likely) and may result in a slight overestimate of the number of beta decays (and resulting neutrinos, which in any event are subdominant with respect to those from pion decay). In photodisintegration models {\tt -M~3} and {\tt -M~4}, the processes implemented are single nucleon ejection (with the nucleon type chosen at random) and alpha-particle ejection.

The model used by default is that by Puget, Stecker and Bredekamp (1976)~\cite{bib:PSBsigma} as refined by Stecker and Salamon (1999)~\cite{bib:SSthresholds}. It does not distinguish between protons and neutrons,
treating only beta-decay stable isobars for each mass number~$2\le A \le 4$ and~$9\le A \le 56$ (51~nuclides in total).
The processes it models are single- and double-nucleon ejection for $\epsilon' < \epsilon_1=30~\MeV$ and multiple-nucleon ejection with tabulated branching ratios for $\epsilon_1 < \epsilon' < \epsilon_{\max}=150~\MeV$.
All other processes, \eg those in which deuterons or alpha particles are ejected,
are neglected in this model. An exception is beryllium-9, for which the only process modelled
is fragmentation into one nucleon and two alpha particles.

If {\tt -M} is used with a non-zero value, the first row of the standard input must contain the number of nuclear species in the model (excluding free nucleons), the energy $\epsilon_1$ and the energy $\epsilon_{\max}$; in the PSB model ({\tt -M 0}) these are 51, $30~\MeV$ and $150~\MeV$ respectively. Then the standard input should include one line for each nuclide, giving cross section parameters as follows:
\begin{center}\begin{tabular}{r|l}
  model & parameter list \\
  \hline
  1, 2 & $Z$, $A$, $\epsilon_1^{\min}$, $\epsilon_2^{\min}$, $\epsilon^0_{1}$, $\xi_1$, $\Delta_1$, $\epsilon^0_{2}$, $\xi_2$, $\Delta_2$, $\zeta$ \\
  3 & $Z$, $A$, $t_N$, $h_N^1$, $x_N^1$, $w_N^1$, $h_N^2$, $x_N^2$, $w_N^2$, $c_N$, $t_\alpha$, $h_\alpha^1$, $x_\alpha^1$, $w_\alpha^1$, $h_\alpha^2$, $x_\alpha^2$, $w_\alpha^2$, $c_\alpha$  \\
  4 & $A$, $Z$, $t_N$, $h_N^1$, $x_N^1$, $w_N^1$, $c_N$, $t_\alpha$, $h_\alpha^1$, $x_\alpha^1$, $w_\alpha^1$, $c_\alpha$ \\
\end{tabular}\end{center}
All energies must be given in MeV and all cross sections must be given in millibarns. The parameter values used with the default model ({\tt -M 0}) are listed in \TableRef{tab:PSBparams}.
The corresponding cross sections are:
\begin{description}
  \item[Models 0 and 1]  
    $$\sigma_i(\epsilon') = \begin{cases}
      \xi_i\frac{\Sigma_\text{d}}{W_i}\exp(-\frac{(\epsilon'-\epsilon^0_i)^2}{\Delta_i^2}),
        & \epsilon_i^{\min}<\epsilon'<\epsilon_1,\quad i=1,2; \\
      b_i\frac{\zeta\Sigma_\text{d}}{\epsilon_{\max}-\epsilon_1}, & \epsilon_1 < \epsilon' < \epsilon_{\max},\quad i=1,2,\ldots;
    \end{cases} $$
    where \begin{align*}
      W_i &= \int_{\epsilon^{\min}_i}^{\epsilon_1} \exp( -\frac{(\epsilon'-\epsilon^0_i)^2}{\Delta_i^2}) \dd{\epsilon'}; &
      \Sigma_\text{d} &= 60\frac{(A-Z)Z}{A}~\MeV~\mb; \end{align*}
    branching ratios $b_i$ from \TableRef{tab:PSBbranch}.
  \item [Model 2]
    $$\sigma_i(\epsilon') = \begin{cases}
      \xi_i \qty(1 + \frac{(\epsilon'-\epsilon^0_i)^2}{\Delta_i^2})^{-1}, & \epsilon_i^{\min}<\epsilon'<\epsilon_1,\quad i=1,2; \\
      \text{same as for models 0 and 1}, & \epsilon_1 < \epsilon' < \epsilon_{\max}. \\
    \end{cases} $$
  \item [Model 3]
    $$\sigma_i(\epsilon') = \begin{cases}
      h_i^1 \qty(1 + \frac{(\epsilon'-x_i^1)^2}{{w_i^1}^2})^{-1} + h_i^2 \qty(1 + \frac{(\epsilon'-x_i^2)^2}{{w_i^2}^2})^{-1},
        & t_i<\epsilon'<\epsilon_1, \qquad i=N,\alpha; \\
      c_i, & \epsilon_1 < \epsilon' < \epsilon_{\max}, \quad i=N,\alpha.
    \end{cases} $$
  \item [Model 4]
      $$\sigma_i(\epsilon') = \begin{cases}
      h_i^1 \exp(-\frac{(\epsilon'-x_i^1)^2}{{w_i^1}}),
        & t_i<\epsilon'<\epsilon_1, \qquad i=N,\alpha; \\
      c_i, & \epsilon_1 < \epsilon' < \epsilon_{\max}, \quad i=N,\alpha.
    \end{cases} $$
\end{description}
For $A \le 4$, PSB cross sections are used, regardless of the photodisintegration model chosen; in the models where the channels implemented are $\sigma_N$ and $\sigma_\alpha$, for these nuclei we use $\sigma_N(\epsilon') = \sum_n n \sigma_n^\text{PSB}(\epsilon')$ and $\sigma_\alpha(\epsilon') = 0$.

The \SimProp~v2r4 package contains the files \nolinkurl{xsect_BreitWigner_TALYS-1.0.txt} and \nolinkurl{xsect_BreitWigner_TALYS-1.6.txt}
 intended for use with {\tt -M 2}, \nolinkurl{xsect_BreitWigner2_TALYS-1.6.txt} intended for use with {\tt -M 3}, and \nolinkurl{xsect_Gauss2_TALYS-restored.txt} intended for use with {\tt -M 4}. The parameter values they contain were found from fits to the
\linebreak
\begin{landscape}
\begin{table}
  \centering
  \scriptsize\begin{tabular}{cc|cccc|cccc|c}
  ~ & ~ & \multicolumn{4}{|c|}{one-nucleon} & \multicolumn{4}{c|}{two-nucleon} & ~ \\
  $A$ & $Z$ & $\epsilon_1^{\min}$ & $\epsilon^0_{1}$ & $\xi_1$ & $\Delta_1$ & $\epsilon_2^{\min}$ & $\epsilon^0_{2}$ & $\xi_2$ & $\Delta_2$ & $\zeta$ \\
  \hline
$56$ & $26$ & $10.2$ & $18$ & $0.98$ & $\phantom{0}8$ & $18.3$ & $22$ & $0.15$ & $\phantom{0}7$ & $0.95$ \\
$55$ & $25$ & $\phantom{0}8.1$ & $18$ & $0.93$ & $\phantom{0}7$ & $17.8$ & $24$ & $0.20$ & $\phantom{0}8$ & $0.95$ \\
$54$ & $26$ & $\phantom{0}8.9$ & $18$ & $0.93$ & $\phantom{0}7$ & $15.4$ & $24$ & $0.20$ & $\phantom{0}8$ & $0.95$ \\
$53$ & $24$ & $\phantom{0}7.9$ & $18$ & $1.03$ & $\phantom{0}7$ & $18.4$ & $24$ & $0.10$ & $\phantom{0}8$ & $0.95$ \\
$52$ & $24$ & $10.5$ & $18$ & $1.08$ & $\phantom{0}7$ & $18.6$ & $24$ & $0.05$ & $\phantom{0}8$ & $0.95$ \\
$51$ & $23$ & $\phantom{0}8.1$ & $19$ & $1.02$ & $\phantom{0}7$ & $19.0$ & $25$ & $0.11$ & $\phantom{0}6$ & $0.95$ \\
$50$ & $24$ & $\phantom{0}9.6$ & $19$ & $1.03$ & $\phantom{0}8$ & $16.3$ & $25$ & $0.10$ & $\phantom{0}6$ & $0.95$ \\
$49$ & $22$ & $\phantom{0}8.1$ & $19$ & $1.03$ & $\phantom{0}8$ & $19.6$ & $25$ & $0.10$ & $\phantom{0}6$ & $0.95$ \\
$48$ & $22$ & $11.4$ & $19$ & $1.03$ & $\phantom{0}8$ & $19.9$ & $25$ & $0.10$ & $\phantom{0}6$ & $0.95$ \\
$47$ & $22$ & $\phantom{0}8.9$ & $19$ & $1.03$ & $\phantom{0}8$ & $18.7$ & $25$ & $0.10$ & $\phantom{0}6$ & $0.95$ \\
$46$ & $22$ & $10.3$ & $19$ & $1.03$ & $\phantom{0}8$ & $17.2$ & $25$ & $0.10$ & $\phantom{0}6$ & $0.95$ \\
$45$ & $21$ & $\phantom{0}6.9$ & $19$ & $0.97$ & $\phantom{0}9$ & $18.0$ & $26$ & $0.15$ & $\phantom{0}8$ & $0.95$ \\
$44$ & $20$ & $11.1$ & $20$ & $0.92$ & $\phantom{0}9$ & $19.1$ & $26$ & $0.20$ & $\phantom{0}8$ & $0.96$ \\
$43$ & $20$ & $\phantom{0}7.9$ & $20$ & $0.97$ & $\phantom{0}8$ & $18.2$ & $26$ & $0.15$ & $\phantom{0}8$ & $0.96$ \\
$42$ & $20$ & $10.3$ & $20$ & $1.02$ & $\phantom{0}7$ & $18.1$ & $26$ & $0.10$ & $\phantom{0}8$ & $0.96$ \\
$41$ & $19$ & $\phantom{0}7.8$ & $20$ & $0.92$ & $\phantom{0}6$ & $17.7$ & $26$ & $0.20$ & $\phantom{0}8$ & $0.96$ \\
$40$ & $20$ & $\phantom{0}8.3$ & $20$ & $0.84$ & $\phantom{0}6$ & $14.7$ & $26$ & $0.28$ & $10$ & $0.96$ \\
$39$ & $19$ & $\phantom{0}6.4$ & $20$ & $0.73$ & $\phantom{0}7$ & $16.6$ & $25$ & $0.38$ & $12$ & $0.98$ \\
$38$ & $18$ & $10.2$ & $18$ & $0.86$ & $\phantom{0}8$ & $18.6$ & $22$ & $0.24$ & $\phantom{0}8$ & $0.98$ \\
$37$ & $17$ & $\phantom{0}8.4$ & $20$ & $0.81$ & $\phantom{0}7$ & $18.3$ & $24$ & $0.28$ & $\phantom{0}7$ & $1.00$ \\
$36$ & $18$ & $\phantom{0}8.5$ & $22$ & $0.82$ & $12$ & $14.9$ & $22$ & $0.25$ & $12$ & $1.00$ \\
$35$ & $17$ & $\phantom{0}6.4$ & $20$ & $0.87$ & $\phantom{0}7$ & $17.3$ & $26$ & $0.22$ & $10$ & $1.00$ \\
$34$ & $16$ & $10.9$ & $22$ & $0.87$ & $12$ & $20.1$ & $22$ & $0.20$ & $12$ & $1.00$ \\
$33$ & $16$ & $\phantom{0}8.6$ & $22$ & $0.82$ & $12$ & $17.5$ & $22$ & $0.25$ & $12$ & $1.00$ \\
$32$ & $16$ & $\phantom{0}8.9$ & $22$ & $0.97$ & $12$ & $16.2$ & $30$ & $0.10$ & $12$ & $1.00$ \\
$31$ & $15$ & $\phantom{0}7.3$ & $21$ & $0.85$ & $\phantom{0}8$ & $17.9$ & $29$ & $0.20$ & $12$ & $1.02$ \\
  \end{tabular}
  \hfil
  \begin{tabular}{cc|cccc|cccc|c}
  ~ & ~ & \multicolumn{4}{|c|}{one-nucleon} & \multicolumn{4}{c|}{two-nucleon} & ~ \\
  $A$ & $Z$ & $\epsilon_1^{\min}$ & $\epsilon^0_{1}$ & $\xi_1$ & $\Delta_1$ & $\epsilon_2^{\min}$ & $\epsilon^0_{2}$ & $\xi_2$ & $\Delta_2$ & $\zeta$ \\
  \hline
$30$ & $14$ & $10.6$ & $20$ & $0.83$ & $\phantom{0}7$ & $19.1$ & $26$ & $0.20$ & $\phantom{0}8$ & $1.04$ \\
$29$ & $14$ & $\phantom{0}8.5$ & $20$ & $0.83$ & $\phantom{0}7$ & $20.1$ & $26$ & $0.20$ & $\phantom{0}8$ & $1.04$ \\
$28$ & $14$ & $11.6$ & $21$ & $1.01$ & $\phantom{0}8$ & $19.9$ & $30$ & $0.02$ & $\phantom{0}8$ & $1.04$ \\
$27$ & $13$ & $\phantom{0}8.3$ & $21$ & $0.80$ & $\phantom{0}8$ & $19.4$ & $29$ & $0.20$ & $12$ & $1.05$ \\
$26$ & $12$ & $11.1$ & $18$ & $0.77$ & $\phantom{0}8$ & $18.4$ & $26$ & $0.20$ & $\phantom{0}8$ & $1.08$ \\
$25$ & $12$ & $\phantom{0}7.3$ & $23$ & $0.77$ & $\phantom{0}9$ & $19.0$ & $28$ & $0.20$ & $\phantom{0}7$ & $1.08$ \\
$24$ & $12$ & $11.7$ & $19$ & $0.94$ & $11$ & $20.5$ & $29$ & $0.03$ & $\phantom{0}6$ & $1.08$ \\
$23$ & $11$ & $\phantom{0}8.8$ & $22$ & $0.83$ & $12$ & $19.2$ & $25$ & $0.12$ & $10$ & $1.09$ \\
$22$ & $10$ & $10.4$ & $22$ & $0.81$ & $12$ & $17.1$ & $21$ & $0.11$ & $\phantom{0}4$ & $1.09$ \\
$21$ & $10$ & $\phantom{0}6.8$ & $22$ & $0.84$ & $12$ & $19.6$ & $25$ & $0.08$ & $\phantom{0}6$ & $1.09$ \\
$20$ & $10$ & $12.8$ & $22$ & $0.87$ & $12$ & $20.8$ & $26$ & $0.05$ & $\phantom{0}8$ & $1.09$ \\
$19$ & $\phantom{0}9$ & $\phantom{0}8.0$ & $23$ & $0.76$ & $14$ & $16.0$ & $29$ & $0.14$ & $14$ & $1.10$ \\
$18$ & $\phantom{0}8$ & $\phantom{0}8.0$ & $24$ & $0.67$ & $\phantom{0}9$ & $12.2$ & $29$ & $0.20$ & $10$ & $1.10$ \\
$17$ & $\phantom{0}8$ & $\phantom{0}4.1$ & $24$ & $0.77$ & $\phantom{0}9$ & $16.3$ & $29$ & $0.20$ & $10$ & $1.10$ \\
$16$ & $\phantom{0}8$ & $12.1$ & $24$ & $0.83$ & $\phantom{0}9$ & $22.3$ & $30$ & $0.04$ & $10$ & $1.10$ \\
$15$ & $\phantom{0}7$ & $10.2$ & $23$ & $0.73$ & $10$ & $18.4$ & $23$ & $0.10$ & $10$ & $1.07$ \\
$14$ & $\phantom{0}7$ & $\phantom{0}7.6$ & $23$ & $0.46$ & $10$ & $12.5$ & $23$ & $0.37$ & $10$ & $1.07$ \\
$13$ & $\phantom{0}6$ & $\phantom{0}4.9$ & $23$ & $0.71$ & $\phantom{0}8$ & $20.9$ & $27$ & $0.05$ & $\phantom{0}8$ & $1.06$ \\
$12$ & $\phantom{0}6$ & $16.0$ & $23$ & $0.76$ & $\phantom{0}6$ & $27.2$ & $27$ & $0.00$ & $\phantom{0}8$ & $1.06$ \\
$11$ & $\phantom{0}5$ & $11.2$ & $26$ & $0.85$ & $11$ & $18.0$ & $26$ & $0.15$ & $11$ & $1.03$ \\
$10$ & $\phantom{0}5$ & $\phantom{0}6.6$ & $25$ & $0.54$ & $11$ & $\phantom{0}8.3$ & $25$ & $0.15$ & $11$ & $1.03$ \\
$\phantom{0}9$ & $\phantom{0}4$ & $\phantom{0}1.7$ & $26$ & $0.67$ & $20$ & $18.9$ & $25$ & $0.00$ & $11$ & $1.00$ \\
$\phantom{0}4$ & $\phantom{0}2$ & $19.8$ & $27$ & $0.47$ & $12$ & $26.1$ & $45$ & $0.11$ & $40$ & $1.11$ \\
$\phantom{0}3$ & $\phantom{0}2$ & $\phantom{0}5.5$ & $13$ & $0.33$ & $18$ & $\phantom{0}7.7$ & $15$ & $0.33$ & $13$ & $1.11$ \\
$\phantom{0}2$ & $\phantom{0}1$ & $\phantom{0}2.2$ & $\phantom{0}5$ & $0.97$ & $\phantom{0}9$ & $\phantom{0}2.2$ & $15$ & $0.00$ & $13$ & $0.00$ \\
\multicolumn{11}{c}{~}
  \end{tabular}
  \caption[PSB cross section parameters]{Parameters of PSB cross sections \cite{bib:PSBsigma} with Stecker and Salamon thresholds \cite{bib:SSthresholds} (all energies in MeV)}
  \label{tab:PSBparams}
\end{table}
\begin{table}
  \centering
  \scriptsize\begin{tabular}{r|ccccccccccccccc}
  $A$ & 1 nuc. & 2 nuc. & 3 nuc. & \multicolumn{11}{l}{etc.}\\
  \hline
  3--4 &   $\phantom{0}80\%$ & $20\%$ & \multicolumn{13}{l}{~}\\
  9 &                $100\%$ & \multicolumn{14}{l}{~} \\
  10--22 & $\phantom{0}10\%$ & $30\%$ & $10\%$ &           $10\%$ & $20\%$ & $20\%\phantom{.0}$ & \multicolumn{9}{l}{~} \\
  23--56 & $\phantom{0}10\%$ & $35\%$ & $10\%$ & $\phantom{0}5\%$ & $15\%$ & $\phantom{0}4.5\%$ &
           $4.0\%$ & $3.5\%$ & $3.0\%$ & $2.5\%$ & $2.0\%$ & $1.8\%$ & $1.5\%$ & $1.2\%$ & $1.0\%$ \\
  \end{tabular}
  \caption[PSB cross section branching ratios]{Branching ratios for the number of nucleons ejected in the PSB model for~$30~\MeV < \epsilon' < 150~\MeV$ as a function of the parent nucleus mass number}
  \label{tab:PSBbranch}
\end{table}
\end{landscape}
~\linebreak
results of TALYS~\cite{bib:TALYS}\index{TALYS}, a program that can simulate nuclear reactions for a variety
of projectile types and a wide range of projectile energies, computing cross sections for all exclusive
channels, $\sigma_{n_\n n_\p n_{\rm d} n_{\rm t} n_{\rm h} n_\alpha}$~being the cross section for the channel
in which $n_\n$~neutrons, $n_\p$~protons, $n_{\rm d}$~deuterons, $n_{\rm t}$~tritium nuclei, $n_{\rm h}$~helium-3 nuclei
and $n_\alpha$~helium-4 nuclei are ejected.

In \nolinkurl{xsect_BreitWigner_TALYS-1.0.txt} and \nolinkurl{xsect_BreitWigner_TALYS-1.6.txt}, the one-nucleon ejection cross section $\sigma_1$ for $\epsilon_1^{\min}<\epsilon'<\epsilon_1=30~\MeV$ was fitted to the sum of TALYS one-neutron and one-proton ejection cross sections $\sigma_{100000} + \sigma_{010000}$; the two-nucleon ejection cross section $\sigma_2$ for $\epsilon_2^{\min}<\epsilon'<\epsilon_1=30~\MeV$ was fitted to the sum of TALYS two-neutron, one-neutron-one-proton, two-proton, and one-deuteron ejection cross sections $\sigma_{200000} + \sigma_{110000} + \sigma_{020000} + \sigma_{001000}$; and the total cross section for $\epsilon_1 < \epsilon' < \epsilon_{\max} = 150~\MeV$ was fitted to the TALYS total cross section, while using the same branching ratios as in the PSB model. The default settings of TALYS-1.0 and TALYS-1.6 were used respectively.

In \nolinkurl{xsect_BreitWigner2_TALYS-1.6.txt} and \nolinkurl{xsect_Gauss2_TALYS-restored.txt},\linebreak the~cross sections for nucleon and alpha-particle ejection $\sigma_N$ and$\sigma_\alpha$ were fitted to
    \begin{align}
        \sigma_N &= \sum_\text{channels} n_N \sigma_{n_\n n_\p n_\mathrm{d} n_\mathrm{t} n_\mathrm{h} n_\alpha} =
        \expval{n_N} \sigma_\text{tot} \text{, where~}n_N = n_\n + n_\p + 2n_\mathrm{d} + 3n_\mathrm{t} + 3n_\mathrm{h}; \label{eq:nucl} \\
        \sigma_\alpha &= \sum_\text{channels} n_\alpha \sigma_{n_\n n_\p n_\mathrm{d} n_\mathrm{t} n_\mathrm{h} n_\alpha}
        = \expval{n_\alpha}\sigma_\text{tot}. \label{eq:alph}
    \end{align}
The rationale for not considering multiple nucleon ejection in these models is that, while the statistical uncertainties associated with the available $\ln A$ measurements
    are in principle small enough to distinguish protons from helium-4, they
    are too large to distinguish consecutive intermediate nuclei, \eg carbon-12
    from carbon-13; a detailed discussion about this is found in the \SimProp~v2r3 documentation \cite{bib:SPv2r3} and in \ReferenceRef{bib:SALpropa}.

As discussed in \ReferenceRef{bib:SALpropa}, in the cases relevant to UHECR propagation, the results from released versions of TALYS used with their default settings have been found to be in much worse agreement with the measured total photodisintegration cross section data
than those from the preliminary version used in \ReferenceRef{bib:Khan}. For this reason, in the file \nolinkurl{xsect_Gauss2_TALYS-restored.txt} the parameters were fitted to results from TALYS-1.6 with settings restored to those used in the preliminary version, as described in~\ReferenceRef{bib:SALpropa}.

Also, all versions of TALYS largely overestimate the cross sections for channels in which alpha particles are emitted for the few nuclides for which measured data for these channels are available; these channels are neglected altogether in the PSB model. (The TALYS-1.6 settings that we restored to those used in the preliminary version only affect the total cross sections but not the branching ratios.)

\section{Flux of secondary neutrinos and gamma rays}\label{sec:appli}
\begin{figure}[t]
        \centering
        \includegraphics[width=0.6\textwidth]{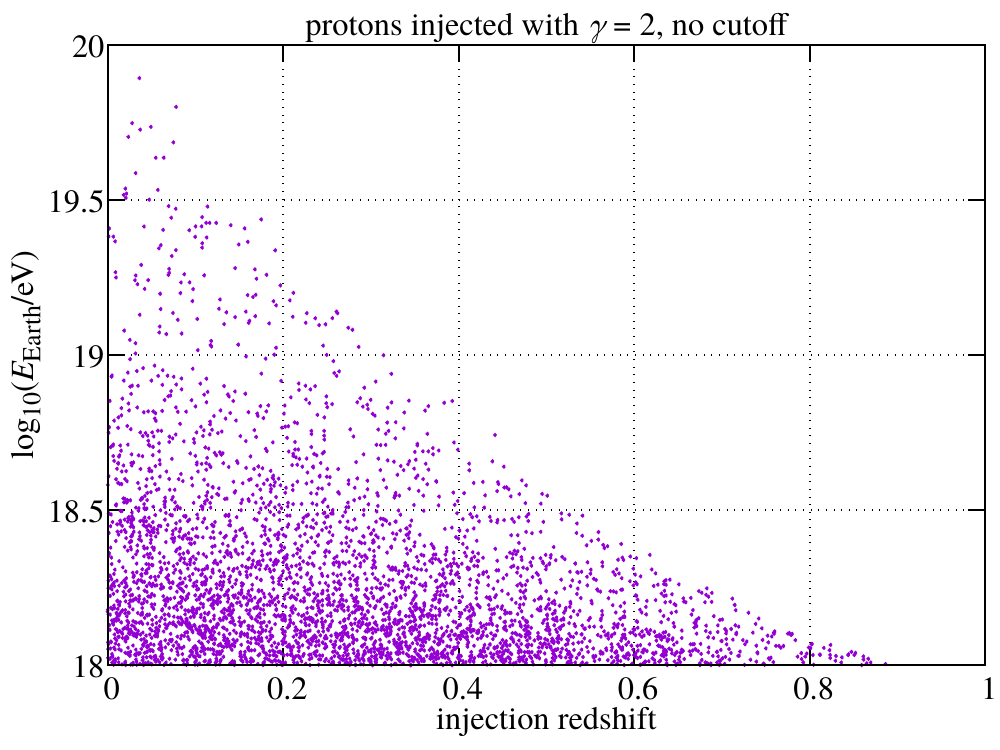}
        \caption{Energies at Earth of protons reaching Earth with $E>1~\EeV$ as a function of their source redshift, showing a horizon at $z\approx 0.9$}
        \label{fig:cutoff}
\end{figure}

The main theoretical interest in secondary neutrinos and gamma rays is that this radiation carries information about sources of UHECRs not otherwise available. For instance, all protons reaching the Earth with $E > 1~\EeV$ must originate from sources at $z < 1$, no matter how much energy they started with (\FigureRef{fig:cutoff}), and this limit is even stricter for other nuclei (especially light ones). Since the flux of extragalactic nuclei with $E \lesssim 1~\EeV$ may be suppressed by diffusion in intergalactic magnetic fields and/or contaminated by the high-energy tail of galactic cosmic rays, this means that all information about sources at $z>1$ is lost when looking at charged particles alone.

On the other hand, neutrinos produced in the decay of (photoproduced) pions can reach the Earth unaffected by their propagation (except for the redshift energy loss and flavour oscillations) even if produced up to $z \sim 10$. Neutrino energies range from a few $\PeV$ (in the case of pion production on EBL photons) to a few $\EeV$ (with CMB photons) \cite{Aloisio:2015ega}. 

Photons from pion decay and electrons/positrons from pair production and pion decay are produced with similar energies and undergo a cascade of pair production ($\gamma_\text{HE} + \gamma_\text{backgr} \to \e^+ + \e^-$) and inverse Compton scattering ($\e^\pm + \gamma_\text{backgr} \to \e^\pm + \gamma_\text{HE}$) with very short interaction lengths, eventually resulting in cascades of photons with up to a few hundred GeV each. 
The shape of the cascade spectrum at Earth does not depend on the energy of the initiating partilce, provided it is sufficiently large $E_\text{prod} \gtrsim 100$~TeV~\cite{bib:secondarygamma}. Therefore, models of UHECR sources which only differ in the source emissivities at large $z$ and would be hard to distinguish from the observation of nuclei alone can still predict very different fluxes of neutrinos and gamma rays \cite{bib:SPv2r2,Heinze:2015hhp,Berezinsky:2016jys}. In principle neutrinos carry more information than the gamma rays, but they are harder to detect, therefore constraints from gamma-ray observations can be more stringent.

Fluxes of secondary gamma rays and neutrinos are also sensitive to the mass composition of UHECRs. For a given energy, heavy nuclei have a lower Lorentz factor and are less likely to undergo pion production with CMB photons. Therefore, the detection of $\EeV$ neutrinos or gamma rays would be evidence for a sizeable fraction of protons among the highest-energy cosmic rays. 

In addition to this, whereas charged particles can be deflected by intergalactic and galactic magnetic fields, neutral particles travel in straight lines so that their arrival direction matches the angular position of the source.

\subsection{Example scenarios}
\begin{figure}[t]
        \centering
        \includegraphics[width=0.60\textwidth]{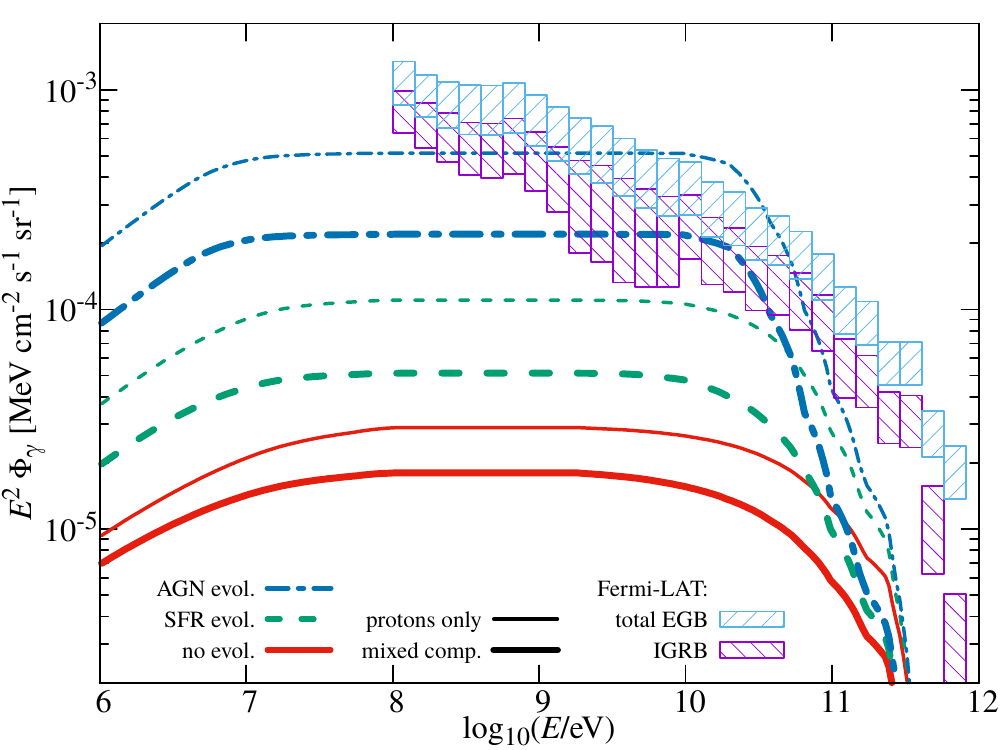}
        \caption{Diffuse extragalactic gamma-ray background expected to result from UHECR propagation in the scenarios of ref.~\cite{Aloisio:2015ega}, assuming various models of source emissivity evolution, using the cascade development model of ref.~\cite{bib:secondarygamma}. Fermi-LAT IGRB and total EGB data~\cite{Ackermann:2014usa} with uncertainties from Galactic foreground modelling are also shown.}
        \label{fig:gamma}
\end{figure}

As an example of the dependence of gamma-ray fluxes on the assumptions about UHECR sources, in figure~\ref{fig:gamma} we show the expected gamma-ray flux from UHECR propagation computed in the three pure-proton scenarios and the three mixed-composition scenarios of ref.~\cite{Aloisio:2015ega}. The EBL model used is {\tt -L 7} (Gilmore \etal 2012 fiducial \cite{bib:GilmoreEBL}). The source emissivities (table~\ref{tab:emissivity}) were normalized to the latest Telescope Array surface detector data \cite{TA-spectrum} in the proton-only scenario and the latest Pierre Auger Observatory combined spectrum data \cite{Auger-spectrum} above $10^{18}~\eV$ in the mixed-composition scenario, while all other parameters were kept fixed to the ref.~\cite{Aloisio:2015ega} values. 
\begin{table}[t]
 \centering
 \begin{tabular}{cccc}
 \hline
  source & proton-only & \multicolumn{2}{c}{mixed-composition scenario} \\
  \cline{3-4}
  evolution & scenario & soft sources & hard sources \\
  \hline
  none & 13.8 & 19.2 & 0.43 \\
  SFR & \phantom{0}6.2 & \phantom{0}8.7 & \\
  AGN & \phantom{0}3.4 & \phantom{0}4.7 &  \\
  \hline
  \multicolumn{4}{c}{all emissivities in $10^{45}~\mathrm{erg}/\Mpc^3/\yr$} 
 \end{tabular}
  \caption{Source emissivities at $z=0$ used in figure \ref{fig:gamma}}
  \label{tab:emissivity}
\end{table}
The gamma-ray spectrum produced by cascading secondary gamma rays or pairs at each given redshift was computed using the analytic approximation of ref.~\cite{bib:secondarygamma} (their eqs.~(9, 10), with~$\mathcal{E}_X$ and~$\mathcal{E}_\gamma$ interpolated as a function of redshift from the values shown in their figures~3 and~4).  
Fermi-LAT measurements~\cite{Ackermann:2014usa} of the isotropic diffuse gamma-ray background (IGRB) and the total extragalactic gamma-ray background (EGB) are also shown for comparison, the difference being the emissions resolved into point sources, which are excluded from the former and included in the latter. Note however that a sizeable fraction of the IGRB may originate in point sources that Fermi-LAT was unable to resolve, making the truly diffuse emission lower~\cite{Liu:2016brs}, but on the other hand if intergalactic magnetic fields are weak enough gamma-rays from UHCER propagation may come from the direction of the original source \cite{Aharonian:2012fu} and be excluded from the IGRB.

The model implemented here, based on the analytic approximation of \cite{bib:secondarygamma}, is very crude, neglecting the fact that the EBL is not monochromatic, and more realistic simulations of the cascade development would give more accurate results. The examples discussed here show how scenarios which would be practically indistinguishable from UHECR observations alone can result in gamma-ray fluxes differing by more than one order of magnitude, allowing observations to discriminate them. Results by other authors with slightly different assumptions \cite{Liu:2016brs,Globus:2017ehu,vanVliet:2017obm} are qualitatively similar, but sometimes more stringent because they added to the predictions (or subtracted from the measurements) expected gamma-ray fluxes from other processes. A more extensive study of this topic will be the subject of a future dedicated publication.

\acknowledgments
We would like to thank CRPropa developers Rafael Alves~Batista, Arjen van~Vliet and David Walz, for helping us determine the origin of the differences between results from early versions of \SimProp and CRPropa, and TALYS developers Arjan Koning and St\'ephane Goriely, for their help in understanding the differences between the various versions of TALYS. We would also like to thank Pasquale Blasi, Carmelo Evoli and Oleg Kalashev for fruitful discussions about the electromagnetic cascades initiated by high-energy electrons and photons in intergalactic photon backgrounds.

The work by DB has received funding from the European
Union's Horizon 2020 research and innovation programme (Grant No. 646623). The work by AdM is supported by the IISN project 4.4502.16.

\appendix
\section{Input parameters and output files}
\label{app:inout}
\SimProp~v2r4 recognizes the following command-line options:
\begin{longtable}[c]{rp{30em}r}
option & description & default \\
\hline\endhead
{\tt -h} & prints the help and exits & none \\
{\tt -s} & seed of the random number generator & 65539 \\
{\tt -N} & number of events to be generated & 100 \\
{\tt -A} & mass number of primary nuclei, $A_\text{inj}$ (chosen at random for each event with~{\tt -A 0}) & 56 \\
{\tt -e} & $\log_{10}(E_{\min}/\eV)$, where $E_{\min}$ is the minimum injection energy & 17 \\
{\tt -E} & $\log_{10}(E_{\max}/\eV)$, where $E_{\max}$ is the maximum injection energy & 21 \\
{\tt -g} & injection spectral index, $\gamma$ & 1 \\
{\tt -z} & minimum source redshift, $z_{\min}$ & 0 \\
{\tt -Z} & maximum source redshift, $z_{\max}$ & 1 \\
{\tt -r} & distance between sources, $L_\text{s}$,\footnote{If $L_\text{s}$ is nonzero, then whenever an event is generated with $\zinj<0.1$ it is rounded up to the next higher integer multiple of $\Delta z =H_0L_\text{s}=L_\text{s}/(4285~\Mpc)$.} in $\Mpc$ & 0 \\
{\tt -L} & EBL model, see \SectionRef{sec:EBL} & 1 \\
{\tt -M} & photodisintegration model, see \SectionRef{sec:disi} & 0 \\
{\tt -n} & nucleon ejection scaling factor (only with {\tt -M~3} and {\tt -M~4}) & 1 \\
{\tt -a} & alpha-particle ejection scaling factor (only with {\tt -M~3} and {\tt -M~4}) & 1 \\
{\tt -D} & beta decay: {\tt 0}~disabled, all nuclei treated as their respective beta-decay stable isobars; {\tt 1}~enabled, treated as instantaneous & 1 \\
{\tt -S} & treatment of pion production: {\tt -1}~continuous energy loss approximation for protons, neglected for other nuclei (as in \SimProp v2r0); {\tt 0}~continuous energy loss for both protons and other nuclei; {\tt 1}~stochastic, on the CMB only; {\tt 2}~stochastic, on both the CMB and the EBL & 1 \\
{\tt -p} & electrons and positrons: {\tt 0}~disregarded, {\tt 1}~individually written to output file (warning: results in very large output files), {\tt 2}~binned according to production redshift (by default $[10^{-4.0}, 10^{-3.8}), \ldots, [10^{+0.8}, 10^{+1.0})$, but can be changed in the file {\tt src/Output.h} before compiling), {\tt 3}~only total energy written & 0 \\
{\tt -o} & output type (see below): {\tt 0}~old ({\tt nuc} and {\tt ev} trees); {\tt 1}~new ({\tt summary} tree); {\tt 2}~both & 0 \\
\end{longtable}

The output is written in a ROOT file, whose name encodes the values of the parameters used (\eg \nolinkurl{SimProp-v2r4_N100_A56_e17.0_E21.0_g1.00_z0.00_Z1.00_r0.00_L1_M0_n1.00_a1.00_D1_S1_p0_o0_s65539.root} when the default settings are used). The file also contains the input parameter values in a tree called {\tt opt}.

If the parameter~{\tt -o} is set to 0, the file contains the following trees:
\begin{description}
\item[Tree {\tt nuc}] This tree has an entry for each branch. The name of the tree is due to historical reasons; particles other than nuclei are also included now.
\begin{longtable}[c]{lp{30em}}
branch & description \\
\hline\endhead
{\tt evt~~~~~~~~~~~~} & event number (starting from 0) \\
{\tt branch} & branch generation number: $0$~for the primary; incremented by~$1$ from the parent branch in stochastic interactions and decays\\
{\tt intmult} & $0$~if the particle reaches Earth; $1$~if the particle undergoes pion production or disintegration with one-nucleon ejection; $n$~if it undergoes $n$-nucleon ejection; $4$~if it undergoes alpha-particle ejection; $1000$~for photons and electrons, whose propagation is not implemented in \SimProp and can be simulated with external tools; $1000+n$~if the particle decays into $n$ particles \\
{\tt Acurr} & mass number during the branch (0 for leptons, photons and pions) \\
{\tt Zecurr} & electric charge during the branch \\
{\tt Flav} & flavours of leptons ($+1$~for $\e^-$ and $\nu_\e$, $-1$~for $\e^+$ and $\bar\nu_\e$, $+2$~for $\mu^-$ and $\nu_\mu$, $-2$~for $\mu^+$ and $\bar\nu_\mu$) at production, $0$~for all other particles\\
{\tt zOri} & redshift at the beginning of the branch \\
{\tt zEnd} & redshift at the end of the branch \\
{\tt EOri} & energy at the beginning of the branch, in~$\eV$ \\
{\tt EEnd} & energy at the end of the branch, in~$\eV$ \\
{\tt Dist} & comoving distance travelled (see \AppendixRef{app:distances}), in~$\Mpc$ \\
\end{longtable}
\item[Tree {\tt ev}] This tree has an entry for each event.
\begin{longtable}[c]{lp{30em}}
branch & description \\
\hline\endhead
{\tt timexev~~~~~~~~} & CPU time used during the event, in seconds \\
{\tt branxev} & total number of branches in the event \\
{\tt seed} & seed of the random number generator at the end of the event \\
\end{longtable}
\end{description}

If the parameter~{\tt -o} is set to 1, the file contains the following tree:
\begin{description}
\item[Tree {\tt summary}] This tree has an entry for each event.
\begin{longtable}[c]{lp{30em}}
branch & description \\
\hline\endhead
{\tt event} & event number (starting from 0) \\
{\tt injEnergy} & injection energy, in $\eV$ \\
{\tt injRedshift} & source redshift \\
{\tt injDist} & source comoving distance (see \AppendixRef{app:distances}), in~$\Mpc$ \\
{\tt injA} & injection mass number \\
{\tt injZ} & injection atomic number \\
{\tt nNuc} & number of nuclei reaching Earth \\
{\tt nucEnergy[nNuc]} & energies of nuclei reaching Earth, in $\eV$ \\
{\tt nucA[nNuc]} & mass numbers of nuclei reaching Earth \\
{\tt nucZ[nNuc]} & atomic numbers of nuclei reaching Earth \\
{\tt nPho} & number of photons produced \\
{\tt phoEProd[nPho]} & production energies of photons, in $\eV$ \\
{\tt phozProd[nPho]} & redshifts of production points of photons \\
{\tt nEle} & when using {\tt -p 0}: 0 (electrons and positrons disregarded);
    \newline when using {\tt -p 1}: number of electrons and positrons produced;
    \newline when using {\tt -p 2}: number of $\e^\pm$ production redshift bins used;
    \newline when using {\tt -p 3}: 1 (only the totals are given)\\
{\tt eleZ[nEle]} & $-1$ for electrons, $+1$ for positrons
 (with {\tt -p 2}: sums over $z_\text{prod}$ bins;    with {\tt -p 3}: sum over all $\e^\pm$)\\
{\tt eleEProd[nEle]} & production energies of $\e^\pm$, in $\eV$
(with {\tt -p 2}: sums over $z_\text{prod}$ bins;   with {\tt -p 3}: sum of $E_\text{prod}/(1+z_\text{prod})$ over all $\e^\pm$)\\
{\tt elezProd[nEle]} & redshifts of production points of $\e^\pm$ 
              (with {\tt -p 2}:  logarithmic centres of $z_\text{prod}$ bins;  with {\tt -p 3}: 0) \\
{\tt nNeu} & number of neutrinos reaching Earth \\
{\tt neuEnergy[nNeu]} & energies of neutrinos reaching Earth, in $\eV$ \\
{\tt neuFlav[nNeu]} & flavours of neutrinos at production ($+1$~for~$\nu_\e$, $-1$~for~$\bar\nu_\e$,
$+2$~for~$\nu_\mu$, $-2$~for~$\bar\nu_\mu$) \\
\end{longtable}
\end{description}

If the parameter~{\tt -o} is set to 2, the file contains all the trees described above.

\section{Distance measures}\label{app:distances}
In an expanding universe, there are several possible definitions of distance which for sizeable~$z$ are not equivalent. The following definitions are valid for flat space ($\Omega = \Om+\OL = 1$).

The comoving distance is the proper distance between the positions of two objects measured at a fixed time, divided by the scale factor~$a(t)=R(t)/R_0=(1+z)^{-1}$ at that time. The comoving distance of two objects moving with the Hubble flow does not vary with time. The comoving distance of an object whose light reaches us today after leaving the object at redshift~$z$ is given by \begin{equation}d_\text{C}(z)= \int_0^z \frac{\dd{z}}{H_0\sqrt{(1+z)^3\Om+\OL}}.\end{equation} Likewise, the comoving volume is given by the proper volume times~$(1+z)^3$; the number density per unit comoving volume of a fixed number of objects moving with the Hubble flow does not vary with time.

The light travel distance is the cosmological time elapsed since the light leaves an object until it reaches us. It is given by \begin{equation}d_\text{T}(z)=\int_0^z \abs{\dv{t}{z}} \dd{z} = \int_0^z \frac{\dd{z}}{H_0(1+z)\sqrt{(1+z)^3\Om+\OL}}.\end{equation}

In the case of a distribution of closely spaced identical sources, we can define the source emissivity~$\mathcal{L}$ as the total energy injected per unit comoving volume per unit time, \ie $\mathcal{L} = n_\text{s} L$, where $n_\text{s}$~is the number density of sources per unit comoving volume, and the luminosity~$L$ of each source is the total energy emitted by the source per unit time,
\begin{equation}
  L = \int_{0}^{+\infty} E Q(E_\text{inj}) \dd{E_\text{inj}},
\end{equation}
$Q$~being the injection spectrum (number of particles emitted per unit energy per unit time) of each source.
Likewise, we can define~$\mathcal{Q}(E_\text{inj})=n_\text{s} Q(E_\text{inj})$.

In the cases where the injection spectrum and density of sources depends on energy and possibly time (or equivalently redshift) but not on position, \eg $\mathcal{Q} = \mathcal{Q}(E_\text{inj},t)$, it can be shown that the expected fluxes at Earth are given by
\begin{equation}
J_i(E_\text{Earth}) =\frac{c}{4\pi} \int_{t_{\min}}^{t_0} \int_0^{+\infty} T_{ij}(E_\text{Earth}|\Einj, t) \mathcal{Q}_j(\Einj,t) \dd{\Einj} \dd{t},
\end{equation}
where $T_{ij}(E_\text{Earth}|\Einj, t)$ is the average number of particles of type~$i$ with energy~$E_\text{Earth}$ at the present time~$t_0$ from each particle of type~$j$ injected with energy~$\Einj$ at time~$t$. Therefore, in the case~$\mathcal{Q}(E_\text{inj},t) = \mathcal{Q}_0(E_\text{inj})S(z)$, in order to correctly analyse simulations of UHECR propagation in which the source positions are sampled from a uniform distribution in~$z$, each event must be weighed by a factor proportional to~$S(z)\abs{\dd{t}/\dd{z}}$.

\section{Performances}
\label{app:time}
In table~\ref{tab:time}, we list the CPU times used by \SimProp~v2r4 simulations in a few example configurations
on machines with AMD Opteron~6238 CPUs with 6 cores and 2.6~GHz clock rates.
The times mainly depend on the
choice of settings about the production of secondary neutrinos, electrons and
photons: producing none ({\tt -S 0 -p 0}) is usually two orders of magnitude faster than
producing all, binning electrons and photons by redshift ({\tt -S 2 -p 2}). 
All other things being equal, the computation time is very roughly proportional to
the injection mass number and redshift, whereas the injection energy does not
have a major impact.
\begin{table}[t]
  \centering
  \begin{tabular}{cc||rrr|rrr||}
    $\zinj$ & $\log_{10}(\Einj/\eV) \to$ & \multicolumn{3}{|c|}{$[18, 19)$} & \multicolumn{3}{|c||}{$[19, 20)$} \\
    $\downarrow$ & $\Ainj\to$ & $1$ & $14$ & $56$ & $1$ & $14$ & $56$ \\
    \hline \hline
    \multirow{4}{*}{$[0.0, 0.2)$}
    & {\tt -S 0 -p 0} & 0.06 & 5.92 & 13.70 & 0.05 & 6.29 & 18.08 \\
    & {\tt -S 1 -p 0} & 1.13 & 9.88 & 23.13 & 1.05 & 12.12 & 30.84 \\
    & {\tt -S 2 -p 0} & 3.24 & 21.78 & 50.62 & 2.93 & 27.05 & 63.37 \\
    & {\tt -S 2 -p 2} & 13.78 & 126.90 & 444.78 & 13.69 & 181.69 & 574.90 \\
    \hline
    \multirow{4}{*}{$[0.2, 0.5)$}
    & {\tt -S 0 -p 0} & 0.09 & 9.83 & 21.15 & 0.08 & 8.62 & 26.40 \\
    & {\tt -S 1 -p 0} & 2.41 & 22.70 & 53.72 & 2.27 & 26.24 & 70.07 \\
    & {\tt -S 2 -p 0} & 6.49 & 48.32 & 129.47 & 5.55 & 63.66 & 173.37 \\
    & {\tt -S 2 -p 2} & 57.81 & 633.57 & 1134.89 & 55.74 & 873.60 & 1775.10 \\
    \hline
    \multirow{4}{*}{$[0.5, 1.0)$}
    & {\tt -S 0 -p 0} & 0.10 & 16.95 & 29.82 & 0.13 & 10.33 & 32.94 \\
    & {\tt -S 1 -p 0} & 3.25 & 47.39 & 99.87 & 3.89 & 46.48 & 134.69 \\
    & {\tt -S 2 -p 0} & 9.36 & 110.51 & 254.83 & 9.87 & 117.60 & 341.75 \\
    & {\tt -S 2 -p 2} & 97.62 & 1407.66 & 2269.57 & 126.48 & 2109.06 & 3936.28 \\
    \hline
    \hline
    \multicolumn{8}{c}{all times in milliseconds per event}
  \end{tabular}
  \caption{\label{tab:time}Computation times required by \SimProp~v2r4, using the PSB photodisintegration model ({\tt -M 0}) and the Gilmore \etal 2012 fiducial EBL model ({\tt -L 7}).  Similar results are obtained with the TALYS photodisintegration model ({\tt -M 4 < xsect\_Gauss2\_TALYS-restored.txt}) and/or the Dom\'inguez \etal 2011 best-fit EBL model ({\tt -L 4}).}
\end{table}


\providecommand{\href}[2]{#2}\begingroup\raggedright\endgroup

\end{document}